\documentclass[sigconf,nonacm]{acmart}
\setcopyright{none}
\acmConference{}{}{}
\acmBooktitle{}
\acmYear{}
\copyrightyear{}
\acmDOI{}
\acmISBN{}
\acmPrice{}
\usepackage{xurl}
\usepackage{colortbl}
\usepackage{float}
\usepackage{placeins}
\usepackage{multirow}
\usepackage{tabularx,array}
\definecolor{TableOurs}{HTML}{EFF3F6}
\definecolor{TableGroup}{HTML}{F4F4F4}
\definecolor{TableAux}{HTML}{777777}

\title[RoFiGR]{RoFiGR: Coarse Routing, Fine Refinement via Level-Adaptive Distillation in Multimodal Generative Recommendation}
\author{Ji Dai\texorpdfstring{\textsuperscript{1}}{}}
\author{Zhipeng Wei\texorpdfstring{\textsuperscript{2}}{}}
\author{Zhuang Zhuang\texorpdfstring{\textsuperscript{2}}{}}
\author{Rongfeng Guo\texorpdfstring{\textsuperscript{2}}{}}
\author{Jie Chen\texorpdfstring{\textsuperscript{2}}{}}
\author{Quan Fang\texorpdfstring{\textsuperscript{1}}{}}
\authornote{Corresponding authors: Quan Fang (\nolinkurl{qfang@bupt.edu.cn}) and Fei Pan (\nolinkurl{panfei05@kuaishou.com}).}
\author{Fei Pan\texorpdfstring{\textsuperscript{2}}{}}
\authornotemark[1]
\affiliation{\institution{\textsuperscript{1}Beijing University of Posts and Telecommunications\\\textsuperscript{2}Kuaishou Technology}\city{Beijing}\country{China}}
\email{daiji@bupt.edu.cn}

\hypersetup{pdfauthor={Ji Dai, Zhipeng Wei, Zhuang Zhuang, Rongfeng Guo, Jie Chen, Quan Fang, Fei Pan},pdfkeywords={generative recommendation, multimodal recommendation, semantic identifiers, knowledge distillation}}

\makeatletter
\renewcommand{\@mkauthors@iii}{%
  \gdef\@currentauthors{}\gdef\@currentaffiliation{}%
  \global\setbox\mktitle@bx=\vbox{%
    \unvbox\mktitle@bx\par\medskip\centering
    {\large Ji Dai\textsuperscript{1}, Zhipeng Wei\textsuperscript{2}, Zhuang Zhuang\textsuperscript{2}, Rongfeng Guo\textsuperscript{2},\par
    Jie Chen\textsuperscript{2}, Quan Fang\textsuperscript{1,*}, Fei Pan\textsuperscript{2,*}\par}
    \smallskip
    {\small \textsuperscript{1}Beijing University of Posts and Telecommunications\par
    \textsuperscript{2}Kuaishou Technology, Beijing, China\par
    \nolinkurl{daiji@bupt.edu.cn}\par}\bigskip}}
\makeatother

\begin{document}
\let\balance\relax
\raggedbottom

\begin{abstract}
Generative recommenders predict an item's semantic identifier (SID) one code at a time. Early codes select broad branches, so a routing error can exclude relevant items before fine-grained prediction begins. In multimodal recommendation, these coarse decisions must reflect collaborative preferences as well as item semantics. We introduce RoFiGR, a two-tree recommender that uses collaborative information in both SID construction and generation. It aligns cumulative L1--L2 code vectors with frozen collaborative item embeddings and explicitly encodes item distance and source level in the serialized history. During generation, a sequential teacher supervises legal branches. Fixed coarse-level weights maintain routing supervision, while a bounded adaptive residual allocates additional weight according to teacher--student demand. On Instruments, Arts, and Games, RoFiGR improves all five ranking metrics over the evaluated external baselines. Five-seed paired tests with Holm correction support the main NDCG@10 improvements. Comparisons with alternative collaborative supervision objectives and tokenizers support the proposed design. Matched task and decoding analyses distinguish the effects of training exposure, stream fusion, and candidate budgets.
\end{abstract}
\maketitle
\pagestyle{plain}
\section{Introduction}
\begin{figure}[!t]
\centering
\includegraphics[width=0.87\linewidth]{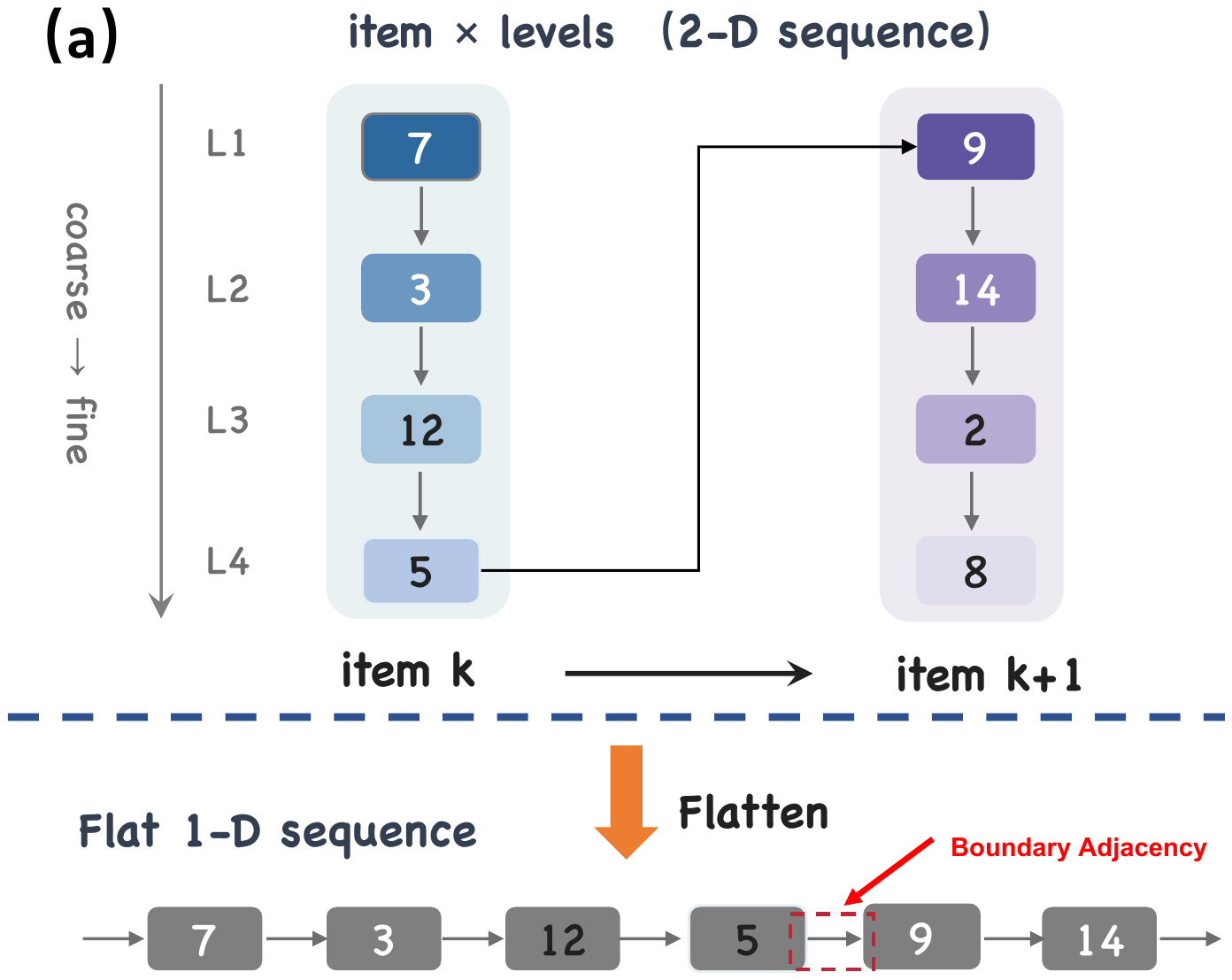}\\[1pt]
\includegraphics[width=0.87\linewidth]{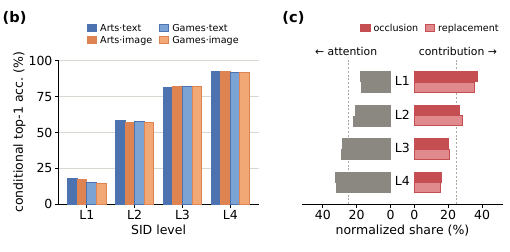}
\caption{(a) Item and level coordinates in a flattened SID history. (b) MACRec conditional accuracy on Arts and Games. (c) Latest-item attention and gold-SID score changes under level-specific occlusion and replacement.}
\label{fig:ab}

\Description{Three panels show flattened semantic identifier sequences, level-wise conditional accuracy, and attention versus prediction-change shares.}
\end{figure}

\begin{figure*}[!t]
  \centering
  \includegraphics[width=0.95\textwidth]{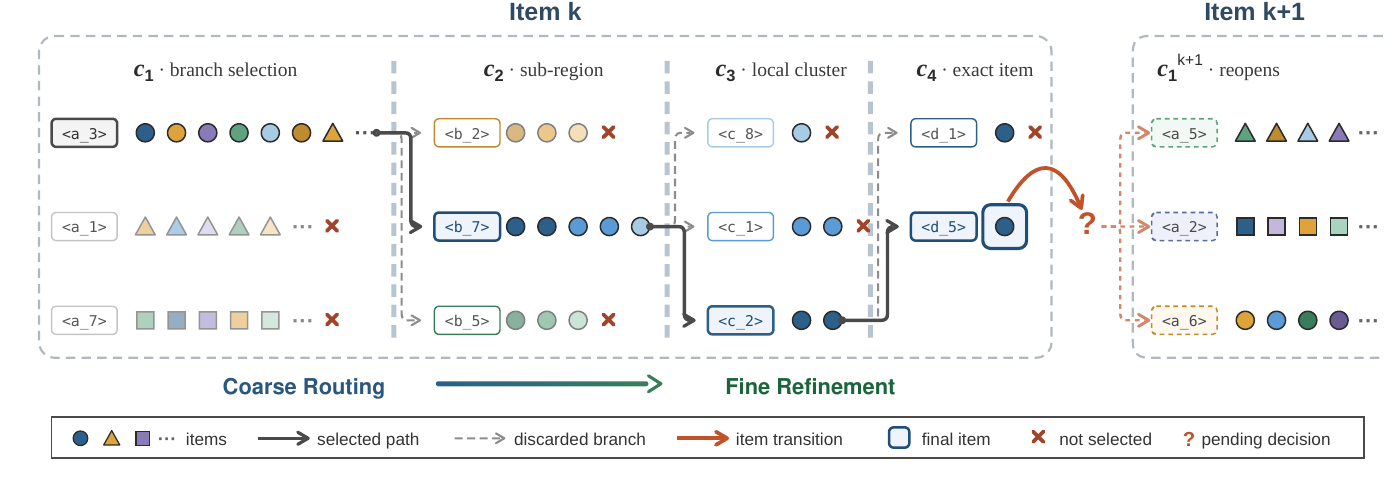}
  \caption{Hierarchical SID generation alternates between coarse routing and
  fine refinement: early codewords select a broad branch, later ones identify
  the item, and each new item restarts routing.}
  \label{fig:rofigr-overview}
\Description{A semantic identifier tree illustrates coarse branch selection followed by fine item refinement and a new routing decision for the next item.}
\end{figure*}

On e-commerce platforms, recommendation draws on both user interactions and product content such as text and images. Sequential models predict the next interaction by scoring candidate items~\citep{sasrec,bert4rec}; generative recommenders instead decode an item's semantic identifier (SID), often learned by residual quantization~\citep{tiger,letter}. Under constrained decoding, early routing decisions determine which products remain reachable within the decoding budget. The challenge is to make these early semantic decisions reflect collaborative preferences while retaining content information for fine-grained product selection. Multimodal approaches incorporate text and image signals into identifiers and generation objectives~\citep{macrec,mmgrec,mql4grec,syngr}; RoFiGR follows MACRec's separate text and image SID trees.

MACRec serializes historical text and image SIDs for a shared encoder--decoder and trains it with token-level cross-entropy. The prediction problem changes along each four-level SID: L1 starts without a predicted prefix, L2 further narrows the branch, and L3--L4 identify the item within it (Figure~\ref{fig:rofigr-overview}). Figure~\ref{fig:ab}(b) shows the level-wise differences in conditional accuracy. A uniform training objective does not explicitly account for these different roles.

Collaborative information can help at two stages. During tokenization, frozen preference-aware item embeddings can align coarse content representations with behavioral similarity. During generation, a sequential teacher can compare the legal branches for a particular user history, providing information that a one-hot next-code label leaves unspecified. The first signal shapes coarse code vectors shared across users; the second guides history-conditioned branch selection.

The history representation presents a related issue. After flattening multiple SIDs, token distance alone does not distinguish item boundaries and source levels. We therefore add item-distance and source-level biases to the first encoder self-attention layer. Figure~\ref{fig:ab}(c) examines these coordinates through attention patterns and prediction changes under level-specific interventions.

RoFiGR combines these components in a two-tree multimodal generator. Fixed coarse-level distillation weights maintain supervision for routing, while a bounded adaptive residual directs additional weight to levels where the teacher provides a useful signal beyond the student and catalog prior. We evaluate alternative collaborative supervision objectives and tokenizers under matched task and decoding budgets, examining how coarse representations and branch supervision contribute to final ranking performance.

Our contributions are: (1) collaborative regularization of cumulative L1--L2 multimodal SID representations, evaluated through both ranking and discrete-tree diagnostics; (2) anchored, demand-adaptive legal-branch distillation with explicit item/source coordinates in history encoding; and (3) controlled comparisons with CF-to-token objectives, CF-aware tokenizers, and matched task and decoding budgets, supported by five-seed paired tests.

\section{Related Work}

\paragraph{Sequential Recommendation.}
Sequential recommenders predict the next interaction from historical behaviors, evolving from recurrent and convolutional encoders~\cite{gru4rec,caser} to self-attentive architectures~\cite{sasrec,bert4rec} enhanced with self-supervised objectives~\cite{s3rec,duorec}. Multimodal variants further exploit item content to enrich preference modeling~\cite{m3srec,mmsr,missrec,hm4sr}. These models score candidate items directly. Their relative item scores can provide soft supervision for a generative recommender.

\paragraph{Generative Recommendation.}
TIGER~\cite{tiger} established the two-stage paradigm that quantizes content embeddings into residual Semantic IDs (SIDs) and autoregressively generates the target identifier. Follow-ups improve identifier quality via collaborative, contrastive, or end-to-end tokenization~\cite{letter,cost,etegrec}. Multimodal extensions broaden what SIDs encode~\cite{mmgrec,mql4grec,musicrec}: MACRec~\cite{macrec} aligns modalities at both tokenization and generation stages, and SynGR~\cite{syngr} promotes cross-modal synergy during generation. RoFiGR focuses on coarse routing in this two-tree setting, combining preference-aware coarse representations with level-adaptive supervision over legal branches.

\paragraph{Hierarchical Structure of Semantic IDs.}
Hierarchical item indexing~\cite{howtoindex} and residual quantization organize item prediction into coarse-to-fine decisions. SID-MLP~\cite{sidmlp} finds that the first token faces the largest candidate space and dominates prediction difficulty; later tokens perform local refinement once the prefix is fixed. How~Well~\cite{howwellgeneralize} identifies a trade-off between compositional generalization through prefix sharing and memorization of specific transitions, a strength of discriminative models~\cite{dig}. Latte~\cite{latte} proves that a fixed semantic tree couples the probabilities of prefix-sharing items regardless of user context. To address such level heterogeneity, GLASS~\cite{glass} aggregates long-term interests for coarse tokens and retrieves local history for fine ones; Hi-SAM~\cite{hisam} assigns cross-modal consensus to coarse levels and modality-specific details to deep ones; APAO~\cite{apao} reinforces vulnerable prefixes; other efforts reorganize level roles in the tokenizer~\cite{cofirec,intrr,resid} or allocate inference computation by position~\cite{s2gr,iba}. RoFiGR uses collaborative information to shape the coarse code space and adapts branch-level supervision to the different demands of SID levels.

\paragraph{Collaborative Token Supervision.}
TCA4Rec \citep{tca4rec} projects collaborative item scores into prefix-conditioned token distributions and optimizes a soft next-token objective. Align$^3$GR \citep{align3gr} studies collaborative--semantic alignment, while MusicRec \citep{musicrec} incorporates collaborative information into multimodal identifiers. Building on item-to-token score aggregation, RoFiGR couples legal-branch supervision with collaborative regularization of cumulative coarse SID vectors, anchored demand-adaptive layer allocation, and relative item/source encoding in a two-tree multimodal generator. Matched controls hold the teacher, student, task exposure, and decoding protocol fixed when comparing CF supervision objectives.
\begin{figure*}[t]
  \centering
  \includegraphics[width=\textwidth]{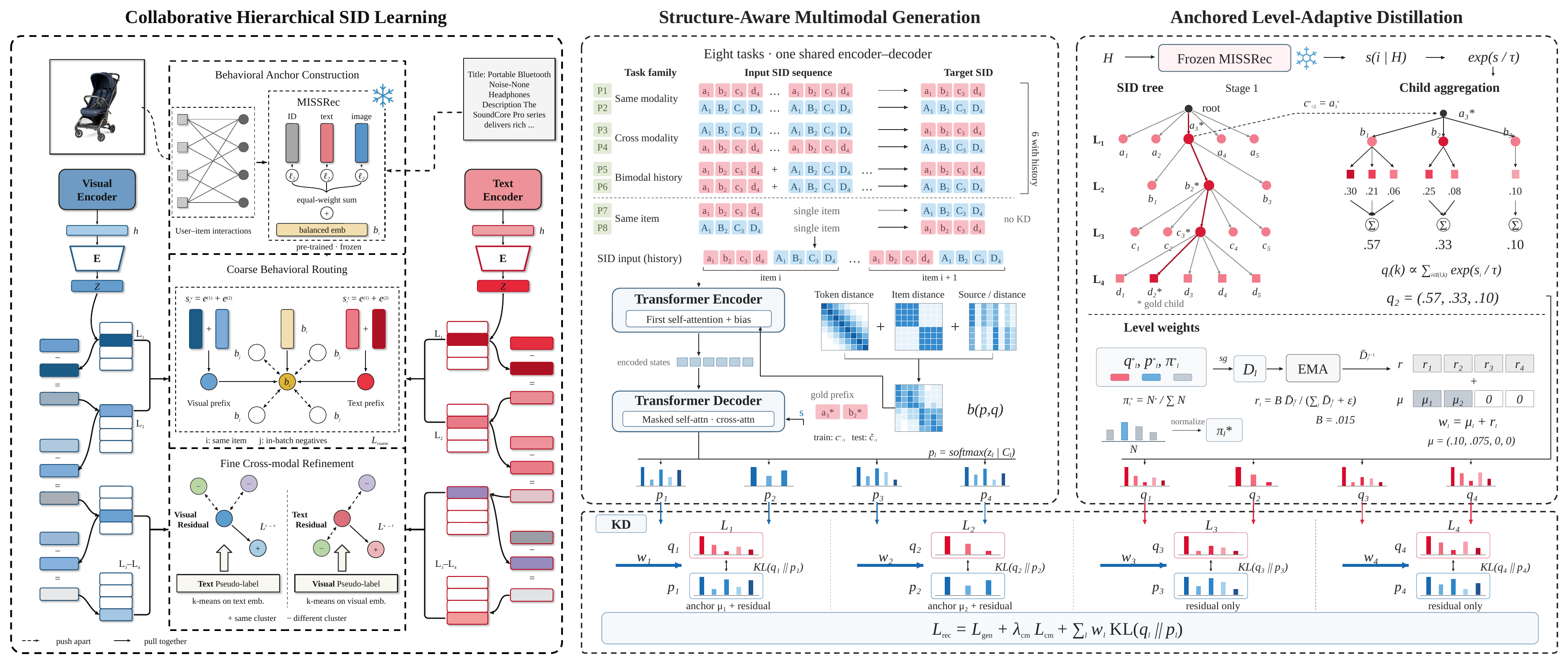}
  \caption{RoFiGR: coarse-code alignment (left), item/source-aware generation (middle), and anchored adaptive distillation (right).}
  \label{fig:main}

\Description{The RoFiGR architecture combines collaborative hierarchical semantic identifier learning, structure-aware multimodal generation, and anchored level-adaptive distillation.}
\end{figure*}

\section{Methods}
\subsection{Preliminaries and Overview}

Given a user history $\mathcal{H}=(i_1,\ldots,i_n)$, the task is to predict the
next item $i_{n+1}$. Each item $i$ has a text or image SID
$\mathbf{c}^{m}(i)=(c^{m}_1(i),\ldots,c^{m}_L(i))$, where
$m\in\{\mathrm{t},\mathrm{v}\}$, $c^{m}_l(i)\in[K_l]$, and $L=4$. An
encoder--decoder $p_\phi$ consumes the serialized SIDs of the history and
generates a target SID autoregressively. Its probability factorizes over the
four levels as $p_\phi(\mathbf{c}^{m}\mid\mathcal{H})=\prod_l
p_\phi(c_l^{m}\mid c_{<l}^{m},\mathcal{H})$.

Let $\mathcal{I}^{m}(c^{m}_{\le l})$ denote the items that share a prefix in
tree $m$. A coarse code therefore selects a set of items, and its probability
aggregates the item probabilities inside that branch:
\begin{equation}
P(c^{m}_{\le l}\mid\mathcal{H})=
\sum_{i\in\mathcal{I}^{m}(c^{m}_{\le l})}P(i\mid\mathcal{H}).
\label{eq:setlevel}
\end{equation}
A frozen MISSRec teacher~\citep{missrec} provides history-independent item embeddings $\mathbf e(i)$ for SID learning and history-conditioned scores $s(i\mid\mathcal H)$ for generator distillation. The static item target is the normalized sum of the separately normalized ID, text, and image representations from the frozen teacher; it is denoted $b_i$ in Figure~\ref{fig:main} and $\mathbf e(i)$ here. We extract these embeddings before tokenizer optimization and precompute scores for the training histories over the effective catalog. Both SID trees are then frozen for generator training. Teacher fitting, scoring, and storage are offline costs; neither generator optimization nor online inference requires teacher execution.

\subsection{Collaborative Hierarchical SID Learning}

We retain MACRec's multimodal residual quantizer~\citep{macrec}, including its
text and image reconstruction, residual quantization, and late cross-modal
interaction.
We align the cumulative L1--L2 code-vector sum with the normalized frozen MISSRec item embedding $\mathbf e(i)$. For modality $m$, the prefix representation
$\mathbf{z}_{\le2}^{m}(i)$ is the sum of the two selected code vectors. We set
$u_{ij}^{m}=\mathrm{sim}(g(\mathbf{z}_{\le2}^{m}(i)),\mathbf{e}(j))/\tau_c$,
where $g$ is a shared linear projector, $\tau_c$ is the alignment temperature, and
$u_{ii}^{m}$ is the positive pair.

For a minibatch $\mathcal{S}$, we use an in-batch contrastive
objective~\citep{simclr} for tree $m$:
\begin{equation}
\mathcal{L}_{\mathrm{coa}}^{m}=-\frac{1}{|\mathcal{S}|}
\sum_{i\in\mathcal{S}}
\log\frac{\exp(u_{ii}^{m})}
{\sum_{j\in\mathcal{S}}\exp(u_{ij}^{m})}.
\label{eq:coarse}
\end{equation}
The final $\mathcal{L}_{\mathrm{coa}}$ averages the text- and image-tree
objectives. The projector is shared by the two trees, whereas their codebooks
remain separate. We apply the alignment once to the cumulative coarse prefix
and leave L3--L4 unconstrained by it. Let $\mathcal{L}_{\mathrm{base}}$ collect
MACRec's reconstruction, quantization, codebook, and fine-level cross-modal
objectives. The tokenizer objective is
$\mathcal{L}_{\mathrm{SID}}=\mathcal{L}_{\mathrm{base}}+
\lambda_{\mathrm{coa}}\mathcal{L}_{\mathrm{coa}}$. Collisions are repaired at L4, or L3--L4 if necessary, without L1--L2 reassignment. Each complete four-token SID maps to one catalog item. We assess the resulting discrete trees through prefix sharing and branch occupancy (Section~\ref{sec:diagnostic}).
Each CF-input, all-level and shuffled-CF tokenizer control learns its own SID mapping. Its trie, occupancy prior, legal teacher-branch targets and generator are rebuilt for that mapping.

\subsection{Structure-Aware Multimodal Generation}
\label{sec:gen}

\paragraph{Multimodal training tasks.}
We retain MACRec's three bidirectional task families: next-item generation,
asymmetric item generation, and item-level identifier translation. Each
family has a text-target and an image-target variant. The first two families
contain a user history, whereas the two translation tasks map the text and
image SIDs of a single item without history. We add a fourth family in which
every historical item supplies both complete identifiers. Its two prompts
generate the next text SID and the next image SID separately. The generator
thus has eight tasks in total, six of which are history-conditioned. All tasks
use independent soft prompts and share the same encoder--decoder.

For a bimodal history, the four text tokens of an item are followed by its
four image tokens, and all eight tokens share one item coordinate. This setup
lets both trees read the same behavior sequence, but flattening still hides
whether two encoder positions belong to the same item. Although level-specific
vocabularies identify each token's own level, the standard positional bias
does not express the source level associated with a pairwise relation.

\paragraph{Structure-aware encoder.}
For query position $p$ and source position $q$, let
$\Delta_{\mathrm{tok}}$ and $\Delta_{\mathrm{it}}$ denote their signed token
and item distances, respectively, and let $l(q)$ be the SID level of the
source token. We clip item distance to $[-19,19]$, which covers the 20-item
history used in our experiments. In the first encoder self-attention layer,
we augment the native token-distance bias with relation-aware
coordinates~\citep{shaw2018}:
\begin{equation}
b(p,q)=b_{\mathrm{tok}}[\Delta_{\mathrm{tok}}]
+b_{\mathrm{it}}[\Delta_{\mathrm{it}}]
+b_{\mathrm{src}}[l(q),\beta(\Delta_{\mathrm{it}})].
\label{eq:bias}
\end{equation}
The source-distance bucket map $\beta$ uses 11 signed bins (Section~\ref{sec:implementation}); item distance uses 39 offsets. The two tables are per-head parameters initialized to zero. The
added biases apply only when both positions correspond to SID tokens and are
set to zero for any pair involving a soft prompt, separator, or padding token. The
item-distance term separates within-item relations from cross-item relations
at the same token distance, while the source-level term distinguishes which
level supplies the information. Together, these biases let encoder attention distinguish item boundaries and source levels while leaving the decoder unchanged.

\subsection{Anchored Level-Adaptive Distillation}\label{sec:distillation}

The generator's one-hot target identifies the observed child but does not say
how the remaining branches compare. Following knowledge and ranking
distillation~\citep{hinton2015,rankingdistill,derrd}, we obtain this missing
information from the frozen teacher. Under the gold prefix
$\mathbf{c}^{m\star}_{<l}$, let
$\mathcal{I}^{m}_{l,k}$ contain the items assigned to legal child $k$, and let
$\mathcal{C}^{m}_l$ denote all legal children. We aggregate the teacher's item
scores into the following softened branch distribution:
\begin{equation}
q_l^{m}(k)=
\frac{\sum_{i\in\mathcal{I}^{m}_{l,k}}
\exp\!\left(s(i\mid\mathcal{H})/\tau_t\right)}
{\sum_{k'\in\mathcal{C}^{m}_l}\sum_{j\in\mathcal{I}^{m}_{l,k'}}
\exp\!\left(s(j\mid\mathcal{H})/\tau_t\right)}.
\label{eq:marg}
\end{equation}
As in CF-to-token projection ~\citep{tca4rec}, this target aggregates teacher item scores over legal children given a gold prefix. We allocate the distillation weight across levels using fixed coarse anchors and a demand-adaptive residual. Text-target tasks use
the text tree and image-target tasks use the image tree. The student
distribution $p_{\phi,l}^{m}$ is restricted to the same legal children and is
conditioned on the gold prefix during training.

The distillation weight combines fixed routing anchors and a bounded residual. For gold child $c_l^\star$, write $q_l^\star,p_l^\star,\pi_l^\star$ for teacher, student, and catalog-count probabilities. Define
\begin{equation}
D_l=\min\{[\log q_l^\star-\log p_l^\star]_+,
[\log q_l^\star-\log\pi_l^\star]_+\}.
\label{eq:demand}
\end{equation}
Here $\pi_l(c\mid u)=N(u,c)/\sum_{c'}N(u,c')$, where $N$ counts catalog items, not interactions. Log arguments are bounded below by $10^{-12}$ only for demand computation; the KL target is unchanged. Demand is positive only when the teacher assigns the gold child more probability than both the student and the catalog prior; the smaller gap limits its magnitude. A single legal child has zero demand. Detached demands are aggregated by valid-example count and update separate tree-specific exponential moving averages. The final level weight is
\begin{equation}
w_l^m=\mu_l+B\,
\frac{(\bar D_l^m)^\gamma}
{\sum_{j=1}^{L}(\bar D_j^m)^\gamma+\epsilon}.
\label{eq:weight}
\end{equation}
We use $\boldsymbol\mu=(0.10,0.075,0,0)$, $B=0.015$, $\gamma=1.5$, $\epsilon=10^{-8}$, and EMA decay $\rho=0.99$. With $S_m=\sum_l(\bar D_l^m)^\gamma$, the realized residual sum is $B S_m/(S_m+\epsilon)$: it starts at zero and approaches the budget $B$ as demand increases. Each tree shares a controller across its three history-conditioned tasks. The two translation tasks receive no history-conditioned KD.

All EMA values start at zero. Each optimizer step uses the previous EMA, aggregates detached demands over accumulated minibatches weighted by valid-example counts, and updates once at the step boundary. Missing-tree steps leave that EMA unchanged. Anchors are active from step one; the residual is disabled for the first 1\% of steps while the EMA updates (80 of 8,000 steps on Instruments; 120 of 12,000 on Arts/Games).

Teacher and student logits use the same gold-prefix legal-child mask before normalization; the default teacher temperature is $\tau_t=1$.

\subsection{Training Objectives and Inference}

RoFiGR is trained in two stages. We first optimize
$\mathcal{L}_{\mathrm{SID}}$ and freeze both materialized SID trees. We then
train the encoder--decoder with the eight generation tasks, the inherited
item-level cross-modal objective, and branch distillation:
\begin{equation}
\mathcal{L}_{\mathrm{rec}}=\mathcal{L}_{\mathrm{gen}}
+\lambda_{\mathrm{cm}}\mathcal{L}_{\mathrm{cm}}
+\sum_{(d,m,l)\in\mathcal{D}}w_l^m
\mathrm{KL}\!\left(q_l^m\,\|\,p_{\phi,l}^m\right).
\label{eq:total}
\end{equation}
Here $\mathcal L_{\mathrm{gen}}$ is cross-entropy across eight tasks, $\mathcal L_{\mathrm{cm}}$ is the inherited item-alignment objective, and $\mathcal D$ contains valid history-conditioned task--tree--level tuples. Each KL term is reduced over valid examples, excluding padding and invalid branches, before applying task and level weights. Matched supervision comparisons use the same history-conditioned task scope, legal-child mask and checkpoint-selection rule. The matched task-exposure controls additionally fix total effective target-token exposure, as detailed in Section~\ref{sec:protocol}. Teacher parameters are frozen; temperatures and objective weights are selected using validation data only.

For a fixed teacher distribution, hard CE plus a weighted branch KL has the same student-dependent term as cross-entropy against a normalized hard--soft mixture, but this identity does not hold for the log-of-mixture soft-NTP objective.

Four streams are decoded independently with trie-constrained beam size 20: \texttt{seqrec}, \texttt{seqimage}, \texttt{biseq2text}, and \texttt{biseq2image}. For decoded item $i$ in candidate set $C_s$,
\begin{equation}
\begin{split}
\ell_s(i)&=\sum_{l=1}^4\log p_s(c_l(i)\mid c_{<l}(i),\mathcal H_s),\\
a_s(i)&=\begin{cases}\exp(\ell_s(i))/\sum_{j\in C_s}\exp(\ell_s(j)),&i\in C_s,\\0,&i\notin C_s,\end{cases}\\
F(i)&=\tfrac14\sum_{s\in\mathrm{all4}}a_s(i).
\end{split}
\label{eq:fusion}
\end{equation}
Full-catalog normalization is used to construct offline teacher targets; online serving uses trie-constrained candidates and stream-local normalization.
The union is mapped to unique item IDs and sorted by $F$. Two-stream controls average only \texttt{seqrec} and \texttt{seqimage}; they retain the same eight-task checkpoint when labeled same2. The two asymmetric history tasks and two same-item translation tasks are training-only. No teacher or reranker is invoked online.

Stream-local normalization puts each candidate set on the same probability scale. Equal averaging then accumulates evidence for items supported by multiple streams, while modality-specific candidates contribute through their originating stream.

\subsection{Implementation}\label{sec:implementation}
Each modality uses four residual codebooks $[64,256,256,512]$ with dimension 32. Collision repair is restricted to L3--L4. A shared projector aligns cumulative L1--L2 vectors with normalized frozen MISSRec embeddings. The T5 generator has four encoder and four decoder layers, hidden size 128, FF size 1,024, and six heads with key/value size 64. The attention projection width is therefore 384. Each task has 16 soft-prompt tokens. Item distances use 39 offsets, clipped to $[-19,19]$; source distance uses 11 bins: $d\le-9$, $[-8,-5]$, $[-4,-3]$, $-2$, $-1$, $0$, $1$, $2$, $[3,4]$, $[5,8]$, $d\ge9$. Added biases contain $6(39+4\cdot11)=498$ scalars and apply only to SID-token pairs in the first encoder layer.

\begin{table*}[t]\centering
\caption{Training tasks, distillation scope, and inference streams.}\label{tab:training_tasks}
\fontsize{8.5}{10.3}\selectfont\setlength{\tabcolsep}{5pt}\renewcommand{\arraystretch}{1.18}
\begin{tabular*}{\textwidth}{@{\extracolsep{\fill}}lllccc@{}}\toprule
Task & Input & Target tree & History & KD & Inference\\\midrule
\rowcolor{TableGroup}\texttt{seqrec} & Text history & Text & Yes & Yes & Yes\\
\rowcolor{TableGroup}\texttt{seqimage} & Image history & Image & Yes & Yes & Yes\\
\texttt{seqimage2item} & Image history & Text & Yes & Yes & ---\\
\texttt{seqitem2image} & Text history & Image & Yes & Yes & ---\\
\rowcolor{TableGroup}\texttt{biseq2text} & Text + image history & Text & Yes & Yes & Yes\\
\rowcolor{TableGroup}\texttt{biseq2image} & Text + image history & Image & Yes & Yes & Yes\\\midrule
\texttt{item2image} & Current text SID & Image & --- & --- & ---\\
\texttt{image2item} & Current image SID & Text & --- & --- & ---\\\bottomrule
\end{tabular*}\end{table*}

\begin{table}[t]
\centering
\caption{Training settings across the three domains.}\label{tab:training_settings}
\fontsize{8.1}{9.6}\selectfont
\setlength{\tabcolsep}{1.4pt}
\begin{tabular}{@{}llccc@{}}
\toprule
Stage & Parameter & Instr. & Arts & Games\\
\midrule
\textbf{Tokenizer} & levels / codebooks & \multicolumn{3}{c}{$4/[64,256,256,512]$}\\
          & code dimension & \multicolumn{3}{c}{32}\\
          & $\lambda_{\mathrm{coa}}$ / $\tau_c$ & \multicolumn{3}{c}{$0.01$ / 0.10}\\
          & optimizer / LR & \multicolumn{3}{c}{Adam / $10^{-3}$}\\
\midrule
\textbf{Generator} & optimizer / schedule & \multicolumn{3}{c}{AdamW / cosine}\\
          & learning rate & \cellcolor{TableOurs}$10^{-3}$ & \cellcolor{TableOurs}$1.25\!\times\!10^{-3}$ & \cellcolor{TableOurs}$1.25\!\times\!10^{-3}$\\
          & matched batch / accumulation & \multicolumn{3}{c}{512 / 2}\\
          & training steps & \cellcolor{TableOurs}8,000 & \cellcolor{TableOurs}12,000 & \cellcolor{TableOurs}12,000\\
          & weight decay / warmup & \multicolumn{3}{c}{$0.01/0.01$}\\
          & $\lambda_{\mathrm{cm}}$ / beam & \multicolumn{3}{c}{$0.01/20$}\\
\midrule
\textbf{Distillation} & $\tau_t$ & \multicolumn{3}{c}{1.0}\\
             & $(\mu_1,\mu_2,\mu_3,\mu_4)$ & \multicolumn{3}{c}{$(0.10,0.075,0,0)$}\\
             & $B/\gamma/\epsilon$ & \multicolumn{3}{c}{$0.015/1.5$/$10^{-8}$}\\
             & EMA decay $\rho$ & \multicolumn{3}{c}{0.99}\\
\bottomrule
\end{tabular}
\end{table}

Table~\ref{tab:training_tasks} lists the task inputs, distillation scope, and serving streams; Table~\ref{tab:training_settings} gives the training settings. Each tree shares one EMA across its three history-conditioned tasks. Teacher probabilities are computed with log-sum-exp under the gold-prefix legal-child mask. EOS is excluded from the four-code item log-score, and ties in final item scores are resolved by ascending item ID.

\begin{table*}[t]
\centering
\caption{Test performance (0--1). Bold: best; underline: best baseline. $\Delta$: relative gain (\%).}\label{tab:overall_results}
\fontsize{9}{10}\selectfont\setlength{\tabcolsep}{2.5pt}\renewcommand{\arraystretch}{1.00}
\begin{tabular*}{\textwidth}{@{\extracolsep{\fill}}l|l|rrrccrrr@{}}\toprule
\textbf{Dataset} & \textbf{Metric} & \textbf{MISSRec} & \textbf{VIP5} & \textbf{TIGER} & \textbf{MQL4GRec} & \textbf{MACRec} & \textbf{SynGR} & \textbf{RoFiGR} & \textbf{$\Delta$ (\%)}\\\midrule
\multirow{5}{*}{Instruments} & HR@1 & 0.0736 & 0.0732 & 0.0763 & 0.0772 & 0.0810 & \underline{0.0894} & \textbf{0.0928} & +3.803\\
 & HR@5 & 0.1105 & 0.0887 & 0.1003 & 0.1080 & 0.1090 & \underline{0.1225} & \textbf{0.1301} & +6.204\\
 & HR@10 & 0.1382 & 0.1068 & 0.1217 & 0.1318 & 0.1321 & \underline{0.1454} & \textbf{0.1583} & +8.872\\
 & NDCG@5 & 0.0900 & 0.0810 & 0.0884 & 0.0917 & 0.0950 & \underline{0.1064} & \textbf{0.1116} & +4.887\\
 & NDCG@10 & 0.0984 & 0.0868 & 0.0952 & 0.0997 & 0.1023 & \underline{0.1137} & \textbf{0.1208} & +6.245\\
\midrule
\multirow{5}{*}{Arts} & HR@1 & 0.0469 & 0.0465 & 0.0543 & 0.0632 & 0.0639 & \underline{0.0715} & \textbf{0.0743} & +3.916\\
 & HR@5 & 0.1021 & 0.0703 & 0.0899 & 0.1003 & 0.1016 & \underline{0.1072} & \textbf{0.1206} & +12.500\\
 & HR@10 & 0.1330 & 0.0955 & 0.1173 & 0.1254 & 0.1293 & \underline{0.1348} & \textbf{0.1565} & +16.098\\
 & NDCG@5 & 0.0710 & 0.0581 & 0.0727 & 0.0821 & 0.0831 & \underline{0.0903} & \textbf{0.0978} & +8.306\\
 & NDCG@10 & 0.0812 & 0.0664 & 0.0815 & 0.0902 & 0.0920 & \underline{0.0992} & \textbf{0.1093} & +10.181\\
\midrule
\multirow{5}{*}{Games} & HR@1 & 0.0202 & 0.0177 & 0.0170 & 0.0202 & 0.0193 & \underline{0.0213} & \textbf{0.0231} & +8.451\\
 & HR@5 & \underline{0.0683} & 0.0480 & 0.0529 & 0.0644 & 0.0675 & 0.0627 & \textbf{0.0758} & +10.981\\
 & HR@10 & 0.1068 & 0.0762 & 0.0870 & 0.1015 & \underline{0.1078} & 0.1008 & \textbf{0.1224} & +13.544\\
 & NDCG@5 & 0.0389 & 0.0329 & 0.0349 & 0.0422 & \underline{0.0434} & 0.0418 & \textbf{0.0493} & +13.594\\
 & NDCG@10 & 0.0507 & 0.0421 & 0.0460 & 0.0541 & \underline{0.0564} & 0.0540 & \textbf{0.0642} & +13.830\\
\bottomrule\end{tabular*}
\end{table*}

\section{Experiments}
\subsection{Experimental Protocol}\label{sec:protocol}
We use Instruments, Arts, and Games from Amazon reviews~\citep{ni2019}. The filtered catalogs contain 6,250, 9,416, and 13,839 items, respectively; the corresponding test populations are 17,112, 22,171, and 42,259 users. The domains follow the MQL4GRec data setting. After 5-core filtering, each user's events are ordered chronologically: the last event is the test target, the preceding event is the validation target, and earlier events form training sequences. Histories contain at most 20 preceding events. Teacher fitting and student optimization use the training partition. The retained event counts are 136,226, 174,079, and 373,514, respectively.

Each user has one held-out test target, evaluated against the full item catalog. For the one-based target rank $R$, HR@$K=\mathbf1[R\le K]$ and NDCG@$K=\mathbf1[R\le K]/\log_2(R+1)$. HR@1/5/10 and NDCG@5/10 use the same per-user rank without sampled negative items. For a generative model, a target absent from the decoded candidate union contributes zero to every reported cutoff metric. Hyperparameters and checkpoints are selected by validation NDCG@10.
We report means over five seeds (42--46); uncertainty is shown as sample SD or paired 95\% confidence intervals where indicated. The overall comparison follows each baseline's stated inference protocol; the controlled analyses use common teacher, tokenizer, task exposure, and decoding settings where relevant. For a seed-paired difference with mean $\bar d$ and sample SD $s_d$, the two-sided 95\% confidence interval is $\bar d\pm t_{0.975,4}s_d/\sqrt5$. Holm correction is applied separately to six primary HR@10/NDCG@10 comparisons, twelve CF-supervision comparisons, twelve tokenizer comparisons, thirty KD-allocation comparisons, and forty-four ablation comparisons. Inference-only variants reuse the corresponding training checkpoint; objective, task, teacher, and tokenizer variants use separately trained students.

\paragraph{Training exposure and decoding budgets.}
The eight tasks are scheduled round-robin with effective batch 1,024 (two accumulated minibatches of 512), for 8,000 optimizer steps on Instruments and 12,000 on Arts/Games. Six-task controls match total target-token exposure. Two streams with beam 20 and four streams with beam 10 both allocate 40 slots; two streams with beam 40 and four streams with beam 20 both allocate 80. The 40-item-cap control first decodes four beams of 20 and selects at most 40 unique items rank-wise round-robin in the order \texttt{seqrec}, \texttt{seqimage}, \texttt{biseq2text}, \texttt{biseq2image}. Each stream follows its SID log-score order and is renormalized over retained items before equal-weight fusion. The cap limits the final candidate union while retaining four-stream decoding cost.

\paragraph{Model selection and hierarchy controls.}
Each anchor is selected from 0.05, 0.075, and 0.10. Residual budgets are 0, 0.015, 0.025, 0.05, and 0.075; the exponent $\gamma$ is selected from 1, 1.5, and 2. Matched soft-target mixing uses $\alpha\in\{0.05,0.10,0.15,0.20\}$; both soft objectives select 0.10 on validation in all domains. Teacher-temperature controls use $\tau_t\in\{0.5,1,2\}$.
The $64/256/256/512$ and $128/128/256/512$ codebook schedules both allow 16,384 theoretical L1--L2 combinations. They use separately learned tokenizers and generators; their two-tree codebooks contain 69,632 and 65,536 scalars. Including generator, prompt, bias, and embedding modules gives 4,825,074 and 4,804,594 parameters, excluding auxiliary projectors and the teacher.

\subsection{Overall Performance}
\begin{table*}[t]
\centering
\caption{Seed-paired improvements over the strongest external baseline for HR@10 and NDCG@10. Differences and intervals are percentage points; Holm correction covers all six comparisons.}\label{tab:primary_comparisons}
\fontsize{8.5}{10.3}\selectfont\setlength{\tabcolsep}{5pt}\renewcommand{\arraystretch}{1.12}
\begin{tabular*}{\textwidth}{@{\extracolsep{\fill}}lllrrrr@{}}\toprule
Domain & Baseline & Metric & Mean difference & SD of difference & 95\% paired CI & Holm $p$\\\midrule
Instruments & SynGR & HR@10 & 1.2950 & 0.1773 & [1.0748, 1.5152] & 0.00016463\\
Instruments & SynGR & NDCG@10 & 0.7158 & 0.2240 & [0.4377, 0.9938] & 0.00202862\\
Arts & SynGR & HR@10 & 2.1668 & 0.0669 & [2.0838, 2.2498] & 0.00000130\\
Arts & SynGR & NDCG@10 & 1.0046 & 0.1059 & [0.8731, 1.1361] & 0.00008757\\
Games & MACRec & HR@10 & 1.4600 & 0.1218 & [1.3088, 1.6113] & 0.00004607\\
Games & MACRec & NDCG@10 & 0.7897 & 0.0599 & [0.7153, 0.8641] & 0.00003948\\
\bottomrule\end{tabular*}
\end{table*}

RoFiGR has the highest mean in all fifteen dataset--metric comparisons (Table~\ref{tab:overall_results}). Averaged across these fifteen comparisons, the relative gain over the strongest external baseline is 9.4274\%. Full RoFiGR uses four streams; the overall MACRec and SynGR baselines retain native two-stream decoding. All six paired HR@10/NDCG@10 comparisons with the strongest external baseline remain significant after Holm adjustment (Table~\ref{tab:primary_comparisons}; Section~\ref{sec:protocol}). Matched controls below separate supervision from differences in tasks and decoding.

\subsection{Ablation and Teacher Controls}\label{sec:ablation}

\begin{figure}[t]\centering
\includegraphics[width=\columnwidth]{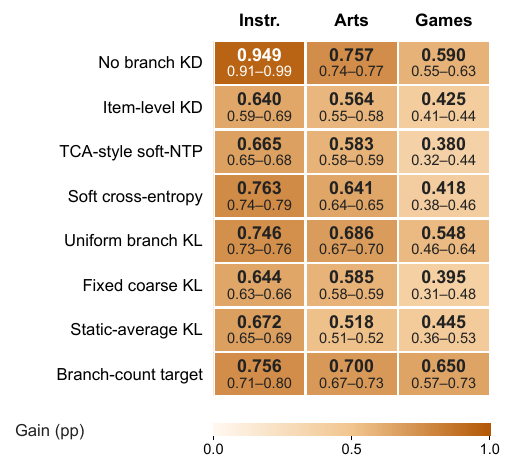}
\caption{NDCG@10 gains over supervision controls (pp; five-seed paired mean, 95\% CI). Colors share one scale.}\label{fig:supervision}
\Description{A matrix compares paired NDCG gains over eight supervision controls across three domains, with confidence intervals in each cell.}
\end{figure}

\begin{figure*}[t]\centering
\includegraphics[width=\textwidth]{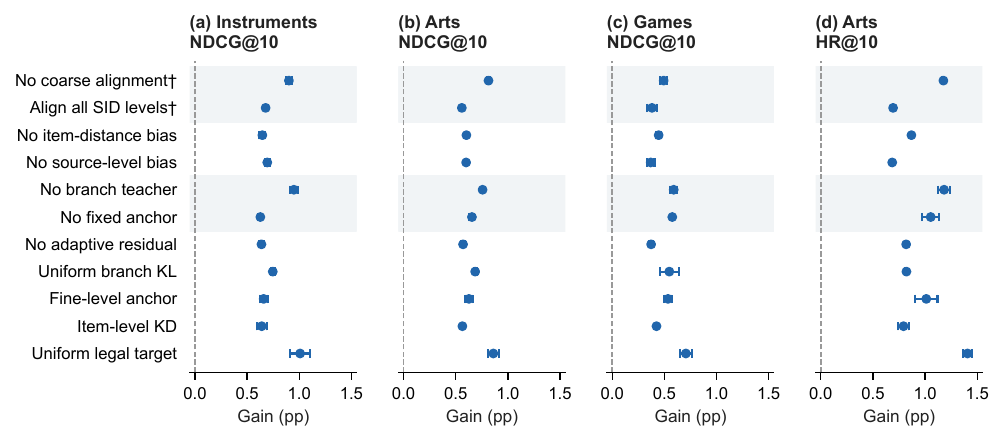}
\caption{Ablation gains (pp; five paired seeds, 95\% CI). $\dagger$: relearned tokenizer. All comparisons pass Holm correction within the 44-test ablation family.}\label{fig:ablation}
\Description{Four panels show paired improvements over eleven ablations, grouped by tokenizer, attention bias and distillation.}
\end{figure*}

Under the common teacher, task-exposure, and four-stream setting, removing coarse SID alignment, item-distance relations, source-level relations, dynamic branch targets, or the adaptive residual consistently lowers mean ranking scores. Figure~\ref{fig:ablation} separates the effects of the SID tokenizer, structural attention, and distillation. The teacher-target controls in Figure~\ref{fig:supervision} further distinguish catalog-item distillation, uniform legal-child supervision, fixed coarse allocation, and adaptive level weighting. Seed-paired NDCG@10 and HR@10 tests support the corresponding improvements after Holm correction, while the aggregate distillation budget is held fixed for nonzero branch-KD controls.

For each tree, nonzero branch-KD allocation controls share the full model's per-step coefficient $W_m(t)$. Uniform allocation uses $(W/4,W/4,W/4,W/4)$, fixed coarse allocation uses $(4W/7,3W/7,0,0)$, and full uses $\boldsymbol\mu+\mathbf r_m(t)$. The no-residual ablation sets $B=0$ and retains $\boldsymbol\mu$, whereas fixed coarse KL follows $W_m(t)$. The uniform legal-child and catalog-count targets assign equal and occupancy-weighted branch probabilities, respectively. Figure~\ref{fig:supervision} reports paired effects for the supervision controls.

\subsection{Hyperparameter Analysis}
\label{sec:hyperparameter}
\begin{figure}[t]
    \centering
    \includegraphics[width=\columnwidth]
    {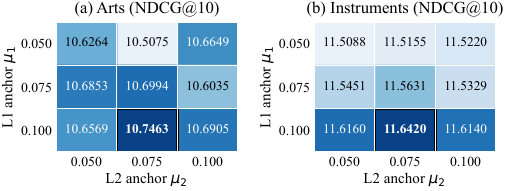}
    \caption{L1/L2 anchor sensitivity: validation NDCG@10 ($\times100$). Outline: selected setting. Colors scale within each dataset.}
    \label{fig:anchor_sensitivity}
\Description{Heatmaps compare validation performance across coarse anchor weights on Arts and Instruments.}
\end{figure}

Within each validation sweep only the named parameter changes; split, task scope, and decoding configuration are held fixed, as are SID trees when applicable. Figure~\ref{fig:anchor_sensitivity} reports all nine anchor pairs for Arts and Instruments. The adopted pair is $(0.10,0.075)$.
\begin{figure}[!t]
    \centering
    \includegraphics[width=\columnwidth]
    {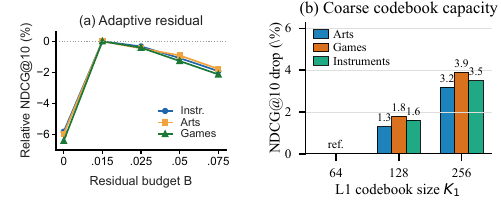}
    \caption{Validation sensitivity to (a) residual budget and (b) L1 capacity. Panel (a) reports $100[N(B)/N(0.015)-1]$; panel (b) retrains each configuration.}
    \label{fig:residual_codebook}

\Description{Plots compare performance for different residual budgets and first-level codebook sizes.}
\end{figure}

The residual sweep uses $B\in\{0,0.015,0.025,0.05,0.075\}$. Setting $B=0$ disables only the residual and retains fixed anchors. The selected budget $B=0.015$ has the best validation mean in all three domains. Removing the residual lowers validation NDCG@10 by 5.82\%, 5.96\% and 6.41\% on Instruments, Arts and Games; larger budgets also reduce performance. The $K_1$ sweep uses separately trained tokenizers and generators. The additional $128/128/256/512$ schedule preserves the codebook-size product; Section~\ref{sec:protocol} gives its parameter counts and the settings for both hierarchy comparisons.

\subsection{Training Tasks and Inference Streams}
\begin{table}[t]
\centering
\caption{Training tasks and inference streams (NDCG@10 $\times100$).}\label{tab:task_streams}
\fontsize{8.1}{9.6}\selectfont\setlength{\tabcolsep}{2.5pt}\renewcommand{\arraystretch}{1.12}
\begin{tabular*}{\columnwidth}{@{\extracolsep{\fill}}lrrr@{}}\toprule
Training / inference & Instr. & Arts & Games\\\midrule
MACRec 6 / native2 & 10.230 & 9.197 & 5.635\\
MACRec 8 / same2 & 10.389 & 9.369 & 5.699\\
MACRec 8 / all4 & 10.588 & 9.591 & 5.821\\
RoFiGR 6 / same2 & 11.314 & 10.279 & 5.961\\
RoFiGR 8 / same2 & 11.454 & 10.368 & 6.050\\
RoFiGR 8 / all4 & 12.085 & 10.928 & 6.425\\
\bottomrule\end{tabular*}
\end{table}

\begin{table}[t]
\centering
\caption{RoFiGR decoding budgets (NDCG@10 $\times100$). All rows reuse the eight-task checkpoint. The cap is applied after decoding.}\label{tab:decoding_budgets}
\fontsize{8.1}{9.6}\selectfont\setlength{\tabcolsep}{2pt}\renewcommand{\arraystretch}{1.12}
\begin{tabular*}{\columnwidth}{@{\extracolsep{\fill}}lrrrr@{}}\toprule
Streams $\times$ beam & Slots & Instr. & Arts & Games\\\midrule
$2\times20$ & 40 & 11.454 & 10.368 & 6.050\\
$4\times10$ & 40 & 11.468 & 10.344 & 6.081\\
$2\times40$ & 80 & 11.476 & 10.331 & 6.059\\
$4\times20$ & 80 & 12.085 & 10.928 & 6.425\\
$4\times20$, cap 40 & 80 & 11.528 & 10.322 & 6.029\\
\bottomrule\end{tabular*}
\end{table}

Table~\ref{tab:task_streams} separates the effects of training tasks and inference streams. With two-stream decoding fixed, eight-task training improves NDCG@10 over six-task training for both MACRec and RoFiGR on all three datasets. Four-stream fusion further improves the same eight-task checkpoints. Table~\ref{tab:decoding_budgets} reports the beam-slot and candidate-cap controls. At 80 beam slots, four-stream decoding improves NDCG@10 over two streams on every domain; at 40 slots, four streams improve Instruments and Games, while Arts is slightly lower. Capping the four-stream union at 40 items reduces its ranking scores on all three domains.

\subsection{Single-Modality Analysis}
\label{sec:single_modality}
\begin{table}[!t]
\centering

\caption{Single-modality students ($\times100$). $\Delta_N$: mean relative gain across domains.}\label{tab:single_modality}
\fontsize{8.1}{9.6}\selectfont
\setlength{\tabcolsep}{1pt}
\renewcommand{\arraystretch}{1.02}
\begin{tabular*}{\columnwidth}{@{\extracolsep{\fill}}lrrrrr@{}}
\toprule
& \multicolumn{2}{c}{Arts}
& \multicolumn{1}{c}{Games}
& \multicolumn{1}{c}{Instr.}
& \multicolumn{1}{c}{\(\bar{\Delta}_N\)} \\
\cmidrule(lr){2-3}
\cmidrule(lr){4-4}
\cmidrule(lr){5-5}
Setting & H@10 & N@10 & N@10 & N@10 & (\%) \\
\midrule
Text-only base & 13.30 & 9.25 & 5.30 & 10.45 & --- \\
+ RoFiGR (text) & 14.15 & 9.85 & 5.66 & 10.93 & +5.96 \\
\midrule
Image-only base & 11.80 & 8.10 & 4.85 & 9.20 & --- \\
+ RoFiGR (image) & 12.55 & 8.55 & 5.15 & 9.70 & +5.73 \\
\bottomrule
\end{tabular*}
\end{table}

Table~\ref{tab:single_modality} compares text-only and image-only students with their corresponding bases under the same input/output modality. Each student is trained independently with one modality and offline supervision from the frozen teacher.

\begin{figure*}[!t]\centering
\includegraphics[width=\textwidth]{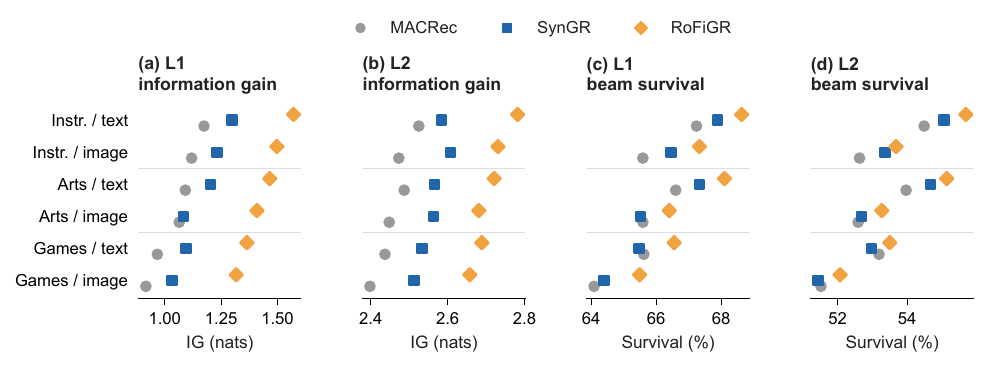}
\caption{Routing by domain and tree: teacher-forced information gain (nats, left) and gold-prefix beam survival (\%, right).}\label{fig:routes}
\Description{Four panels separate information gain and beam survival, with all three models, domains and modality trees.}
\end{figure*}

\subsection{Diagnostic Analysis}
\label{sec:diagnostic}
\begin{table}[t]
\centering
\caption{
Coarse-level diagnostics on Arts (\%; masking drops in pp).
}
\label{tab:diagnostic}

\small
\renewcommand{\arraystretch}{0.98}

\begin{tabular*}{\columnwidth}{
    @{\extracolsep{\fill}}lrrrr@{}
}
\toprule
Method
& \shortstack{L1\\Acc.$\uparrow$}
& \shortstack{L1:2\\Acc.$\uparrow$}
& \shortstack{Err.\\@L1$\downarrow$}
& \shortstack{L1 Drop\\(pp)} \\
\midrule

\multicolumn{5}{@{}l}{\textit{Text SID tree}} \\
MACRec & 17.73 & 10.29 & 89.16 & 3.70 \\
SynGR & 18.95 & 11.27 & 88.68 & 4.20 \\
\rowcolor{TableGroup}
RoFiGR & 22.53 & 14.23 & 87.13 & 5.04 \\
\midrule
\multicolumn{5}{@{}l}{\textit{Image SID tree}} \\
MACRec & 17.26 & 9.83 & 89.38 & 3.37 \\
SynGR & 17.73 & 10.47 & 89.50 & 3.60 \\
\rowcolor{TableGroup}
RoFiGR & 21.55 & 13.49 & 87.85 & 4.97 \\
\bottomrule
\end{tabular*}
\end{table}

\begin{figure*}[t]\centering
\includegraphics[width=\textwidth]{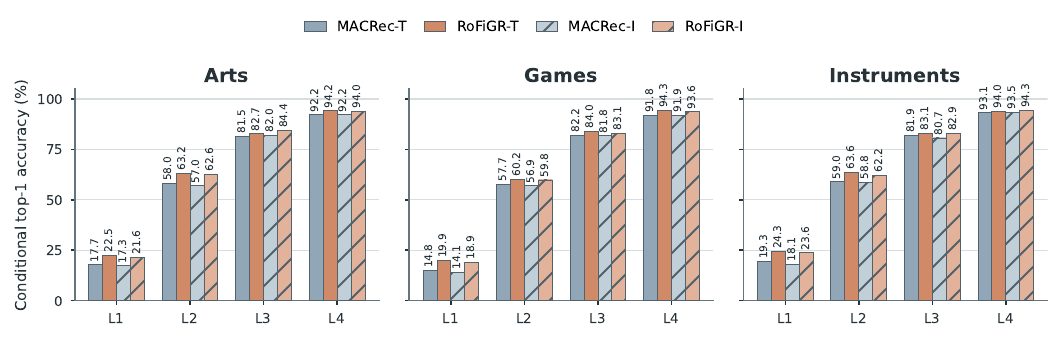}
\caption{Greedy accuracy by SID level, conditioned on correct preceding prefixes.}\label{fig:conditional_accuracy}
\Description{Three-domain conditional-accuracy comparison for MACRec and RoFiGR text and image trees.}
\end{figure*}

Let $N_l$ count examples whose first $l$ greedy SID tokens are correct, with $N_0$ the full test count. Then L1 accuracy is $N_1/N_0$, L1:2 is $N_2/N_0$, and conditional accuracy is $N_l/N_{l-1}$. The first-error share among failed complete SIDs is $(N_0-N_1)/(N_0-N_4)$. These statistics use the same greedy decoding trajectories. Beam survival instead measures whether the gold prefix remains in the beam after each expansion. Figure~\ref{fig:routes} reports information gain and beam survival across domains and trees; Figure~\ref{fig:conditional_accuracy} gives conditional accuracy. Teacher-forced information gain is $\mathrm{IG}_l=\mathbb E[\log p_l-\log\pi_l]$, where $\pi_l(c^*\mid u)=n(u,c^*)/n(u)$ is the catalog-occupancy prior under the gold-prefix legal-child mask. Routing measurements follow the main-comparison protocol and are distinct from final fused item ranks.

On Arts, RoFiGR improves L1 and L1:2 accuracy in both trees (Table~\ref{tab:diagnostic}). L1 Drop measures sensitivity to masking: unmasked minus masked HR@10, in pp. Occlusion zeros the selected level embedding of the latest historical item; replacement uses the same-level embedding from a fixed catalog permutation. Length, item/source coordinates, target SID, and model parameters stay fixed. Prediction change is the absolute change in complete gold-SID log-likelihood, averaged over users, trees, and seeds before normalization over four levels. Attention share is first-encoder attention mass on the latest item's four tokens, normalized within those tokens. The two interventions use the same samples; paired bars show occlusion above replacement on each side of Figure~\ref{fig:ab}(c).

\begin{table}[t]
\centering
\caption{Neighbor sharing (\%). No CA: no coarse alignment; Adj./Eff.: adjusted sharing/effective branches for Full.}\label{tab:partitions}
\fontsize{8.1}{9.6}\selectfont\setlength{\tabcolsep}{2.5pt}\renewcommand{\arraystretch}{1.12}
\begin{tabular*}{\columnwidth}{@{\extracolsep{\fill}}lrrrr@{}}\toprule
Domain / tree & No CA & Full & Adj. & Eff.\\\midrule
Instr. / text & 13.89 & 31.87 & 30.73 & 62.04\\
Instr. / image & 11.75 & 27.17 & 25.96 & 62.31\\
Arts / text & 14.85 & 30.90 & 29.74 & 62.12\\
Arts / image & 11.91 & 27.18 & 25.97 & 62.32\\
Games / text & 13.88 & 31.19 & 30.04 & 62.37\\
Games / image & 11.92 & 27.08 & 25.87 & 62.50\\
\bottomrule\end{tabular*}
\end{table}

\paragraph{Discrete partitions.}
To examine how continuous alignment affects the discrete trees, we measure behavioral-neighbor prefix sharing relative to catalog-occupancy chance, effective L1 branches $\exp(H)$, and empty branches. Across all domains and both trees, RoFiGR has higher neighbor sharing than the no-alignment tokenizer while retaining more than 62 effective branches out of 64 and no empty L1 branches (Table~\ref{tab:partitions}). For percentage-valued sharing, chance adjustment is $100(\mathrm{Share}-\mathrm{Chance})/(100-\mathrm{Chance})$. Behavioral neighbors are constructed from train-only unordered co-occurrences within item-position distance 1--3. With $C_i=\sum_jC_{ij}$ and $E=\sum_iC_i$, candidates are ranked by
\begin{equation}
\left[\log\frac{(C_{ij}+1)E}{(C_i+1)(C_j+1)}\right]\frac{C_{ij}}{C_{ij}+20}.
\end{equation}
We retain at most 20 nonself neighbors per item, breaking ties by ascending item ID. Both trees in each domain use the same directed pair set. For branch occupancy $n_b$, chance sharing is $100\sum_bn_b(n_b-1)/[I(I-1)]$. Each tokenizer is evaluated on its own materialized trie.

\begin{table}[t]
\centering
\caption{NDCG@10 by teacher group ($\times100$). $N$: users; Fixed: budget-matched coarse KL.}\label{tab:teacher_slices}
\fontsize{8.1}{9.6}\selectfont\setlength{\tabcolsep}{2.5pt}\renewcommand{\arraystretch}{1.12}
\begin{tabular*}{\columnwidth}{@{\extracolsep{\fill}}p{105pt}rrr@{}}\toprule
Domain / teacher group & $N$ & Full & Fixed\\\midrule
Instruments / Coarse only & 2330 & 11.264 & 10.744\\
Wrong coarse & 13451 & 12.194 & 11.524\\
Exact item & 1331 & 12.419 & 11.822\\
Arts / Coarse only & 3339 & 11.025 & 10.462\\
Wrong coarse & 17802 & 10.861 & 10.265\\
Exact item & 1030 & 11.780 & 11.323\\
Games / Coarse only & 6409 & 10.869 & 10.131\\
Wrong coarse & 34960 & 4.101 & 3.897\\
Exact item & 890 & 65.719 & 60.294\\
\bottomrule\end{tabular*}
\end{table}

\paragraph{Fixed user slices.}
We group users by predictions from one frozen MISSRec checkpoint and one fixed full-model text tree: exact-item correct, item-wrong but L1--L2 correct, and coarse incorrect. Group membership is fixed across students and seeds. RoFiGR outperforms fixed coarse KL in every group on all three datasets (Table~\ref{tab:teacher_slices}). For the frozen checkpoint used to define these groups, the teacher correctly predicts the exact item for 1,331/17,112, 1,030/22,171, and 890/42,259 users across the three domains.

Behavioral neighbors share prefixes more often while nearly all coarse branches remain active. Improved prefix survival in Figure~\ref{fig:routes} keeps targets reachable for later SID decisions, connecting coarse representation learning with the candidates available for final ranking.

\begin{figure}[!t]\centering
\includegraphics[width=\columnwidth]{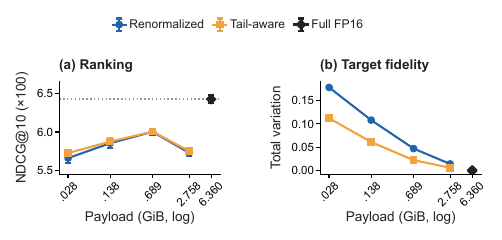}
\caption{Cache trade-offs on Games: student NDCG@10 (five-seed mean $\pm$ SD, left) and teacher-target total variation (right).}\label{fig:teacher_cache}
\Description{Two panels compare cache payload with student ranking and teacher distribution fidelity, including nonmonotonic student results at K=2000.}
\end{figure}

\subsection{Efficiency Analysis}
\label{sec:efficiency}
On Games, total preparation and training take 260 minutes for MACRec and 337 minutes for RoFiGR, including 226 and 241 minutes for generator training and 34 and 38 minutes for SID construction, respectively. RoFiGR adds 58 minutes of teacher preparation. Offline preparation fits the teacher and scores the training histories. Reuse requires the same teacher, split, histories, temperature, and cache representation. Full-prefix FP16 payloads are $2M|\mathcal I|$ bytes, or 0.988, 1.887, and 6.360 GiB across the three domains, before serialization overhead.

With one L20, FP16, batch size one, history length 20, and beam size 20 per stream, RoFiGR's two-/four-stream mean latencies are 31.8/56.7 ms on Instruments, 33.5/60.9 ms on Arts, and 35.1/63.7 ms on Games. Four-stream fusion improves ranking at higher decoding cost; neither mode calls the teacher online. Figure~\ref{fig:teacher_cache} shows the cache trade-off: more faithful teacher targets do not yield monotonic ranking improvements. Timing uses 100 warmup requests followed by 1,000 synchronized requests, including encoding, decoding, item mapping, normalization, fusion, and sorting.

For cache compression, full item logits are computed in FP32 and stored in FP16. Top-$K$ renormalization retains cached items within the gold prefix, falling back to the catalog prior for empty support. Tail-aware caching additionally stores an FP16 log-normalizer, recovers retained mass in FP32, distributes the remaining nonnegative mass uniformly over the omitted $I-K$ items, and normalizes within the prefix. On Games, $M=246,737$ histories and $I=13,839$ items give theoretical payloads $2MI$, $6MK$, and $6MK+2M$ bytes for full, renormalized, and tail-aware caches, excluding serialization metadata. All conditions retain full-catalog teacher scoring.

\subsection{Distinguishing CF Token Supervision from Coarse Routing}\label{sec:cfmatch}

TCA4Rec \citep{tca4rec} projects CF item scores to prefix-conditioned token targets. We compare a TCA4Rec-style soft-NTP matched adaptation, conventional soft cross-entropy, uniform branch KL, fixed coarse KL, and RoFiGR's anchored adaptive KL under identical teacher scores, SID trees, student backbone, task-token exposure, and four-stream inference. Figure~\ref{fig:supervision} shows consistent improvements for RoFiGR across Instruments, Arts, and Games. The paired NDCG@10 differences against the four CF supervision controls are positive and pass Holm correction on all three datasets. Separate CF-input, all-level and shuffled-CF tokenizer controls test coarse-code geometry with the generator architecture and decoding protocol fixed. The CF-input control mixes content and collaborative embeddings with weights 0.8 and 0.2 before projection to 32 dimensions; every tokenizer variant rebuilds its SID mappings, trie, occupancy prior, and teacher targets.

For teacher target $q$, student $p$ and gold child $k^*$, let $y_k=(1-\alpha)\mathbf1[k=k^*]+\alpha q_k$. The two matched soft objectives are
\begin{align}
\mathcal L_{\rm soft\text{-}NTP}&=-\log\sum_k y_kp_k,\\
\mathcal L_{\rm soft\text{-}CE}&=-\sum_k y_k\log p_k.
\end{align}
Soft-NTP adapts TCA4Rec~\citep{tca4rec}; soft-CE uses cross-entropy with the same target. Both replace hard CE only on six history tasks. The mixture coefficient $\alpha$ controls the soft-target weight, whereas $W$ scales the added KL loss. The no-KD and item-level KL controls test the presence and granularity of teacher supervision. Uniform, fixed-coarse and adaptive branch KL retain hard CE on all eight tasks and allocate distillation over the six history tasks as $(W/4,W/4,W/4,W/4)$, $(4W/7,3W/7,0,0)$, and $\boldsymbol\mu+\mathbf r_m(t)$, respectively. All retain the cross-modal loss.

Fixed-tree controls evaluate history-conditioned branch selection; tokenizer controls evaluate item grouping. Gains under matched candidate budgets support the contribution of both decisions to final ranking.

\section{Conclusion}
RoFiGR combines collaborative coarse SID learning with history-conditioned branch supervision in a two-tree multimodal recommender. Coarse-code alignment shapes the candidate groups, item/source-aware attention preserves the structure of the interaction history, and anchored adaptive distillation guides early branch selection. Experiments on Instruments, Arts, and Games show consistent ranking gains, supported by tokenizer, supervision, and candidate-budget controls. The diagnostics connect these gains to greater behavioral-neighbor sharing and improved prefix survival. Together, the results identify coarse routing as a useful point for bringing collaborative preferences into generative recommendation. The two- and four-stream configurations also provide a practical accuracy--latency choice with offline teacher supervision.

\FloatBarrier
\begingroup
\emergencystretch=3em
\sloppy
\bibliographystyle{ACM-Reference-Format}
\bibliography{references}
\endgroup
\end{document}